\documentclass[a4paper]{jpconf}

\usepackage{iopams}
\usepackage{hyperref}
\usepackage{aas_macros}
\usepackage{xspace}
\usepackage{graphicx}
\usepackage{amsmath}

\newcommand{\gws}{gravitational waves\xspace}
\newcommand{\gw}{gravitational wave\xspace}

\begin{document}

\title{A new code for parameter estimation in searches for \gws from known
pulsars}

\author{M~Pitkin$^1$, C~Gill$^1$, J~Veitch$^{2,3}$, E~Macdonald$^1$ and
G~Woan$^1$}

\address{$^1$ SUPA, School of Physics \& Astronomy, University of Glasgow,
Glasgow, G12 8QQ, UK}
\eads{\mailto{matthew.pitkin@glasgow.ac.uk},
\mailto{colingill.glasgow@gmail.com}, \mailto{e.p.macdonald.glasgow@gmail.com},
\mailto{graham.woan@glasgow.ac.uk}}

\address{$^2$ School of Physics \& Astronomy, Cardiff University, Queens
Buildings, The Parade, Cardiff, CF24 3AA, UK}
\address{$^3$ Nikhef -- National Institute for Subatomic Physics, Science Park
105, 1098 XG Amsterdam, The Netherlands}
\ead{johnv@nikhef.nl}

\begin{abstract}
We describe the consistency testing of a new code for \gw signal parameter
estimation in known pulsar searches. The code uses an implementation of nested
sampling to explore the likelihood volume. Using fake signals and simulated 
noise we compare this to a previous code that calculated the signal parameter
posterior distributions on both a grid and using a crude Markov chain Monte
Carlo (MCMC) method. We define a new parameterisation of two orientation angles
of neutron stars used in the signal model (the initial phase and polarisation
angle), which breaks a degeneracy between them and allows more efficient
exploration of those parameters. Finally, we briefly describe potential areas
for further study and the uses of this code in the future.
\end{abstract}

\section{Introduction}\label{sec:intro}

Over the last decade several searches have looked for \gws from a
large selection of known pulsars using data from the LIGO, GEO\,600 and Virgo
\gw detectors \cite{Abbott:2004, Abbott:2005, Abbott:2007, Abbott:2008,
Abbott:2010, Abadie:2011}. No evidence for signals has been observed, but these
searches have produced upper limits on the \gw amplitude for many pulsars by
using Bayesian inference to calculate posterior probability distributions on the
unknown signal parameters \cite{Dupuis:2005}. These searches have all assumed a
single signal model in which the pulsars are triaxial and emit at precisely
twice their rotational frequency, with the complex heterodyned and narrow-banded
time domain signal described by
\begin{align}\label{eq:complexsignal}
h(t) = & h_0\left(\frac{1}{4}F_+(t,
\psi)(1+\cos{}^2\iota)\cos{\phi_0} +
\frac{1}{2}F_{\times}(t, \psi)\cos{\iota}\sin{\phi_0} \right) + \nonumber \\
&  i h_0 \left(\frac{1}{4}F_+(t, \psi)(1+\cos{}^2\iota)\sin{\phi_0} -
\frac{1}{2}F_{\times}(t, \psi)\cos{\iota}\cos{\phi_0} \right),
\end{align}
where the four unknown parameters are the \gw amplitude, $h_0$, polarisation
angle, $\psi$, cosine of the inclination angle, $\cos{\iota}$ and the initial
phase, $\phi_0$, and $F_+$ and $F_{\times}$ are the detector antenna patterns
for the '+' and '$\times$' polarisations for the pulsar's known sky position.

In early searches the posterior probability distribution was calculated on a
4-dimensional grid and the marginalisation integral required to produce
posteriors on individual parameters was performed numerically using the
trapezium rule. Later searches, which began allowing for inclusion of narrow
uncertainties on extra signal parameters (such as frequency and sky position),
performed the posterior exploration using a very basic Markov chain Monte Carlo
(MCMC) method \cite{Abbott:2010}. Both these methods, as applied in the search
code, present practical implementation issues that mean they are not very
flexible to all situations we may encounter: the former grid-based method
quickly becomes computationally unworkable for larger parameters spaces, wastes
time on posterior volumes with tiny probabilities and requires tuning of the
grid for different strength signals; the MCMC method implemented was crude,
required extensive tuning for certain signals and was not able to produce a
Bayesian evidence value for model comparison\footnote{Note that the problems
with the MCMC method that was implemented are not intrinsic issues of all MCMC
methods, but were a problem with the particular basic approach used. Indeed
MCMC outputs can even be used to compute Bayesian evidences
(e.g.~\cite{Weinberg:2009}).} (which could in the future be used as a detection
statistic as in \cite{Prix:2009}).

Recently a new package of codes for Bayesian inference and evidence
calculation ({\tt lalinference}) have been created within the gravitational wave
software repository {\tt lalsuite} \cite{lalsuite}. This suite contains more
sophisticated MCMC methods (e.g.~\cite{Sluys:2008, Roever:2007}) and a method of
Bayesian evidence calculation called nested sampling \cite{Veitch:2010} that are
able to more effectively explore posteriors without detailed tuning, can handle
many parameters and provide values of the Bayesian evidence. We have used this
new suite, and in particular its implementation of nested sampling (which can
provide posterior probability distribution samples as well as an evidence value)
to create a new code for the known pulsar search. Both new and old codes use the
same likelihood as defined in \cite{Dupuis:2005}, which assumes Gaussian noise
with an its unknown variance analytically marginalised over.

This paper has two objectives: i) to show consistency between the new code and
the previous code that has been used in astrophysical searches, and ii) to
introduce a new parameterisation of two of the expected signal's angular
parameters, which breaks a degeneracy that leads to a bimodal posterior
distribution. Finally, we briefly discuss how this new code will be used in
future known pulsar searches.

\section{The algorithm}

Bayesian inference is underpinned by Bayes theorem, which for a problem defined
by a set of variables $\vec{\theta}$, a dataset $d$ and some prior
assumptions $I$, gives the posterior probability density
\begin{equation}\label{eq:bayes}
p(\vec{\theta}|d, I) = \frac{p(d|\vec{\theta}, I) p(\vec{\theta})}
{p(d|I)},
\end{equation}
where $p(d|\vec{\theta}, I)$ is the likelihood function, $p(\vec{\theta})$
is the prior probability distribution of the variables, and $p(d|I)$ is a
normalising factor, often called the Bayesian evidence or the {\it marginalised
likelihood}, which makes the total probability unity. The Bayesian evidence
value can be very useful in testing different hypotheses on the same dataset
i.e.\ models with different variables. It can be found by integrating, or {\it
marginalising}, the numerator of \eref{eq:bayes} over the prior ranges of all
variables
\begin{equation}\label{eq:evidence}
Z = p(d,I) = \int_{\vec{\theta}} p(d|\vec{\theta}, I) p(\vec{\theta}) {\rm
d}\vec{\theta}.
\end{equation}
For many problems this integral will not be analytical, so numerical
integration is required, which will become computationally unfeasible if
calculating the integral on a grid for a large parameter space.

Nested sampling was developed \cite{Skilling:2006} as an efficient method of
performing the integral given by \eref{eq:evidence} even for large parameter
spaces. For good descriptions of the algorithm see \cite{Skilling:2006,
Veitch:2010, Sivia:2006}. As well as producing the evidence the routine will
also output points sampled from the prior parameter volume and their associated
likelihoods. These collected samples can be used to reconstruct the posterior
probability distributions of the parameters.

\section{Change of parameters}

\begin{figure}[!htbp]
\begin{center}
\includegraphics[width=1.0\textwidth]{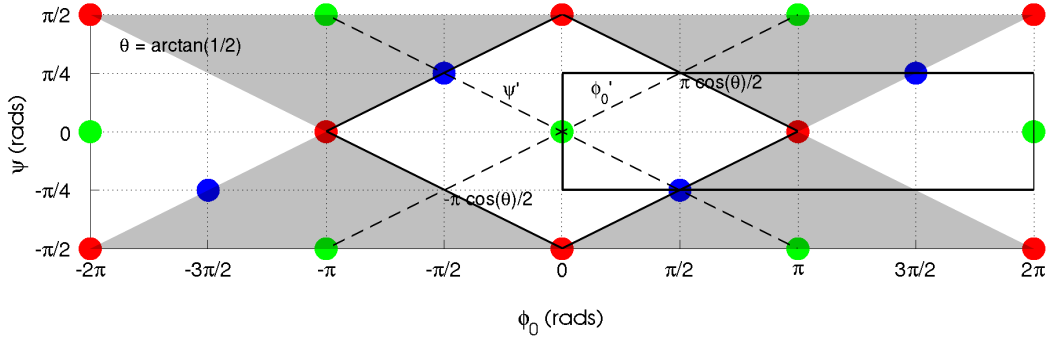}
\caption{A diagram of the $\phi_0$--$\psi$ parameter space demonstrating the new
${\phi'}_0$--$\psi'$ parameterisation. The filled circles represent an area of
probability for a signal, with the same colours representing the same signal
and showing the degeneracy between parameters. The rectangular box shows the
parameters space used in the original code and the diamond box shows the new
parameter space.}\label{fig:phipsi}
\end{center}
\end{figure}

Other than the major change in the underlying algorithm used, a notable change
that has been made in the new code is how it samples two of the signal's angle
parameters: the initial phase $\phi_0$ and the polarisation angle $\psi$. It is
well known that these two parameters are degenerate to a $\pi/2$\,radians shift
in $\psi$ and $\pi$\,radians shift in $\phi_0$ (see Figure~\ref{fig:phipsi}).
This issue can cause problems when trying to efficiently sample this parameter
space. Within the nested sampling algorithm new samples are drawn by generating
an MCMC with a proposal distribution calculated from the covariance of a set of
previously sampled points. For a parameter that is circular about a particular
prior range (e.g.\ an angle parameter wrapping round from $0$ to $2\pi$)
circularity can be easily taken into account when calculating the covariance,
but the degeneracy in the two mentioned parameters can lead to a bimodal
distribution in $\phi_0$. This can be seen in Figure~\ref{fig:phipsi}, for
example with the signal represented by the blue circles. This bimodality will
give a vastly distorted covariance matrix and lead to very low acceptance rates
of new points within the MCMC sampling.

A way to avoid this is to find a new parameterisation in which both parameters
are circular without causing any discontinuity and subsequent bimodality. Such a
parameterisation is given by
\begin{align}
{\phi'}_0 &= \phi_0\sin{\theta} + \psi\cos{\theta}, \nonumber \\
\psi' &= -\phi_0\sin{\theta} + \psi\cos{\theta},
\end{align}
where $\theta = \arctan{(1/2)}$. These new axes are shown by the dashed black
lines in Figure~\ref{fig:phipsi}. It can be seen that signals lying at the edges
of this parameter space wrap around cleanly  (the blue and red circles) at the
edge of the ranges of $-(\pi/2)\cos{\theta} \leq {\phi'}_0, \psi' \leq
(\pi/2)\cos{\theta}$.

Given flat priors on $\phi_0$ and $\psi$, the above change in variables will
mean that the priors on ${\phi'}_0$ and $\psi'$ remain flat (the Jacobian is
constant). The new parameters can be converted back to $\phi_0$ and $\psi$ via
inversion of the transformation matrix giving
\begin{align}
\phi_0 &= {\phi'}_0/(2\sin{\theta}) - \psi'/(2\sin{\theta}), \nonumber \\
\psi &= {\phi'}_0/(2\cos{\theta}) + \psi'/(2\cos{\theta}).
\end{align}
However, it should be noted that this will return values in the range $-\pi
\leq \phi_0 \leq \pi$ and $-\pi/2 \leq \psi \leq \pi/2$, so these have to be
appropriately wrapped to cover the standard range.

\begin{figure}[!htbp]
\begin{center}
\includegraphics[width=1.0\textwidth]{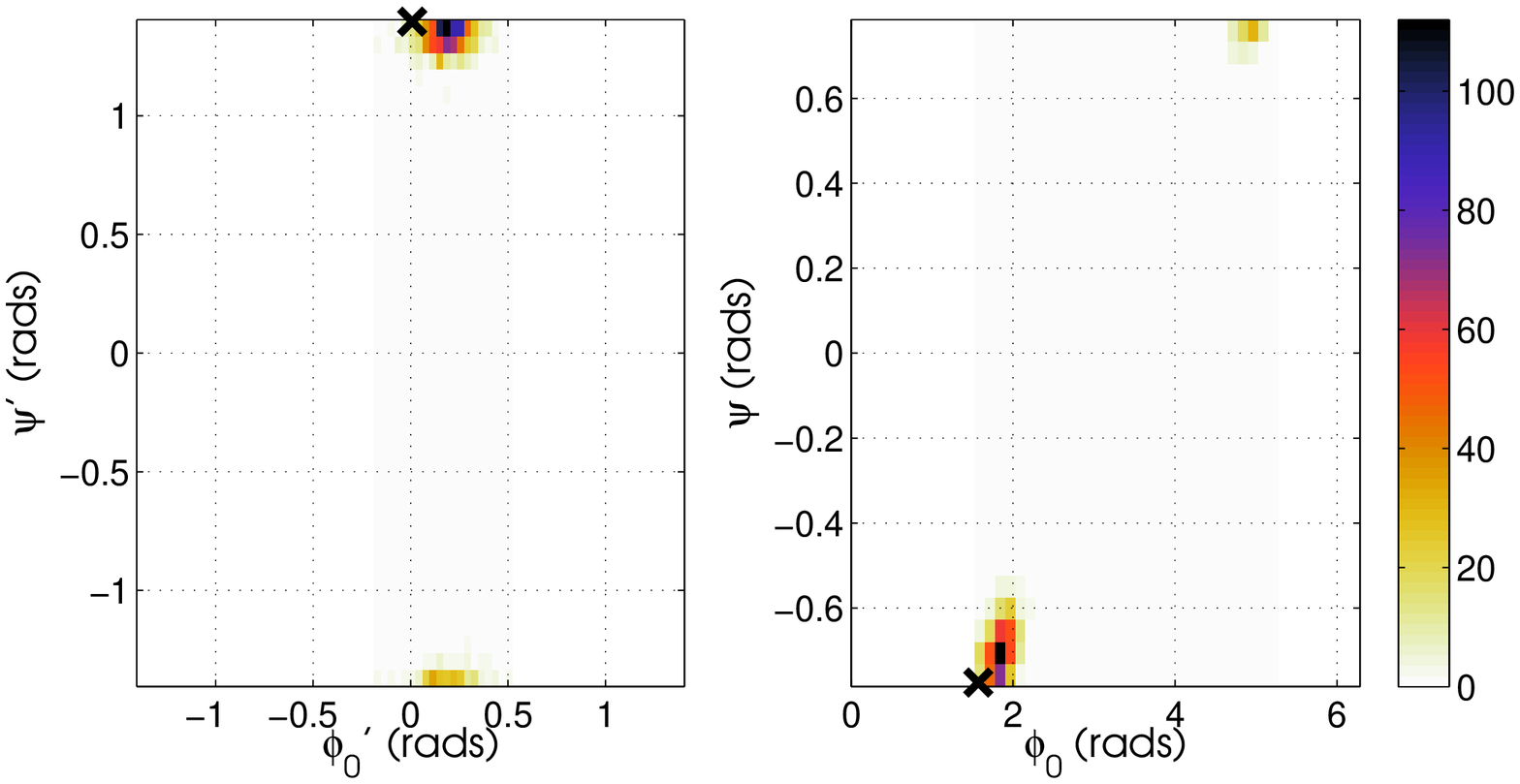}
\caption{The 2D posterior probability distributions of a fake signal (true
parameters marked with the $\times$) extracted with the new
${\phi'}_0$--$\psi'$ parameterisation (left) and converted back into the
$\phi_0$--$\psi$ parameters (right).}\label{fig:phipsi_real}
\end{center}
\end{figure}

An example of this new parameterisation on the calculation of a signal's
parameter posteriors can be seen in Figure~\ref{fig:phipsi_real}, where a strong
fake signal was created at the $\psi$ boundary, with $\phi_0 = \pi/2$\,rads and
$\psi = -\pi/4$\,rads. It was recovered using the new parameterisation and then
converted back into the original parameters correctly, where the bimodality in
$\phi_0$ can clearly be seen. The MCMC acceptance rate improved from a fraction
of a percent to of order 50\%. The same result could also be achieved by using
this new parameterisation to define a prior volume for $\phi_0$ and $\psi$.

\section{Code testing}

The old code used in the analyses up to now has been sufficient, but as stated
previously has some disadvantages. However, as it is well understood it provides
a good check against which to test that the new code is working correctly. We
have tested the posterior probability distributions extracted with simulated
signals and simulated noise using the new and old codes.

\begin{figure}[!htbp]
\begin{center}
\includegraphics[width=0.9\textwidth]{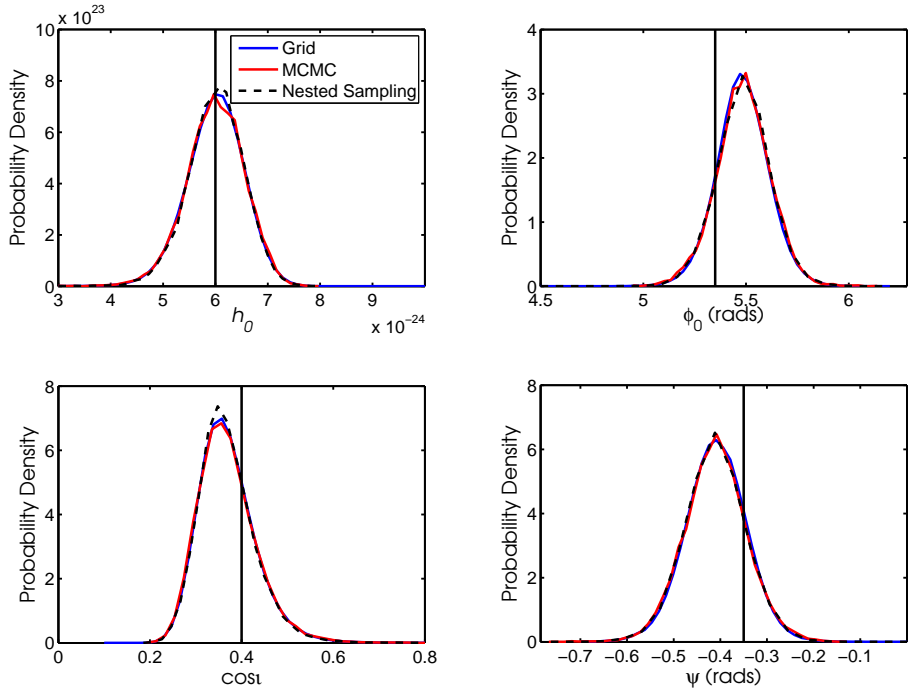}
\caption{Marginalised posterior probability densities for a fake pulsar signal
injected into simulated LIGO Hanford detector noise. The injection
parameters are shown as the thin vertical lines. The solid blue lines are
the posteriors from the old code with a grid-based approach, the solid red lines
are the posteriors from the old code with the basic MCMC approach, and the
dashed black lines are from the new code using nested
sampling.}\label{fig:injection_H1}
\end{center}
\end{figure}

An example is shown in Figure~\ref{fig:injection_H1}, which shows three sets of
posterior probability distributions for the four unknown signal parameters of a
simulated pulsar signal injected into fake noise from the LIGO Hanford (H1)
detector. Two were calculated with the old code in a grid-based mode (with a
$50^4$ grid over the parameter ranges) and an MCMC mode (with a burn-in of
50\,000 points and 100\,000 points sampling the posterior) and the third was
produced using the new code with the nested sampling algorithm (using a total
of 1\,250\,000 likelihood evaluations). Here we make no assessment of the
relative speed/efficiencies of the methods, which mainly depends on the number
of likelihood evaluations they each perform, but just test their consistency. As
with other injections we have studied all three methods show very good agreement
in the posteriors they produce. It should be noted that in setting up the codes
the nested sampling code required no tuning, whereas with the old code for the
grid-based search and the MCMC the parameter ranges and proposal distributions
had to be specified to quite closely match the posterior ranges. All three
methods also recover the true signal values within the posteriors.

\subsection{Upper limit testing}

\begin{figure}[!htbp]
\begin{center}
\includegraphics[width=0.7\textwidth]{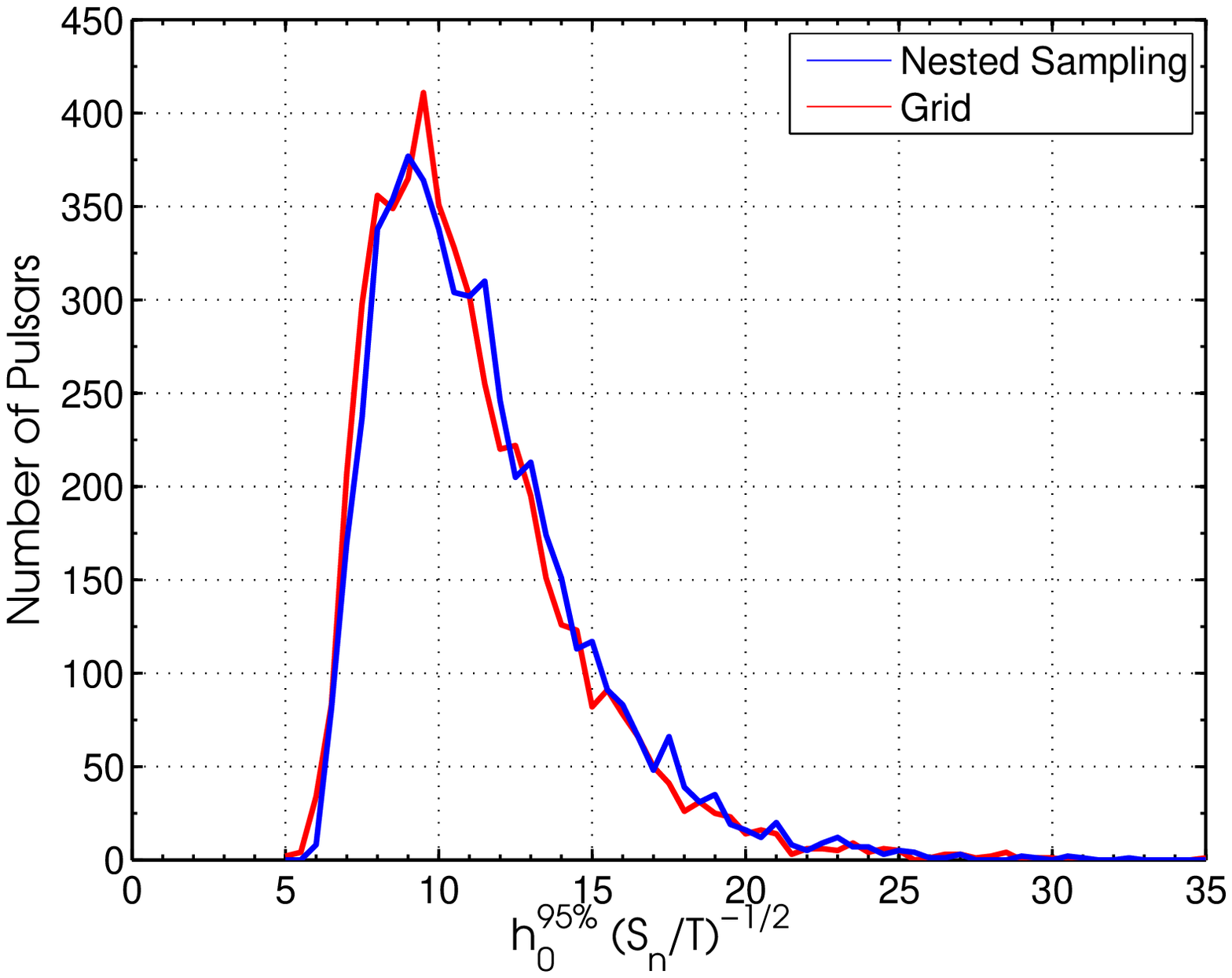}
\caption{The distribution of 95\% degree-of-belief upper limits on $h_0$ for
5000 realisations of noise for sources uniformly distributed on the sky using
the nested sampling code (blue line) and grid-based code (red line).}
\label{fig:ul_distribution}
\end{center}
\end{figure}

Previous astrophysical searches have not seen any signals, so have instead set
upper limits on the \gw amplitude, $h_0$. Monte Carlo tests have shown the
distribution of upper limits (in our case limits at 95\% degree-of-belief) that
would be expected given many datasets consisting of Gaussian noise of known
power spectral density for pulsars randomly distributed on the sky
\cite{Dupuis:2005}. So, again a good way to test the new code is to check that
it can produce the same distribution of upper limits as that found with the old
code. Figure~\ref{fig:ul_distribution} shows the distribution of 95\% upper
limits normalised by the noise power spectral density ($S_n$ in Hz$^{-1}$)
and data length ($T$ in seconds) using the nested sampling code and the old
grid-based code.
These two independent methods are very consistent with each other. The agreement
between these distributions and that shown in \cite{Dupuis:2005} is good, but
not prefect. Without knowing the exact set-up of that earlier analysis we
cannot track down the root cause of this discrepancy, but are confident that
our current analysis with these two independent methods is correct.

\section{Additional models and parameters}

The simple 4-dimensional parameter space of the signal given in
\eref{eq:complexsignal} can easily be covered on a grid (although as stated
still requires some tuning of grid spacing and parameter ranges). However, there
are many cases where the number of parameters in known pulsar searches can
greatly increase and model comparisons are required. Here we briefly discuss a
couple of cases of this and plans for the future.

\subsection{Inclusion of phase parameter uncertainties}
\begin{figure}[!htbp]
\begin{center}
\includegraphics[width=1.0\textwidth]{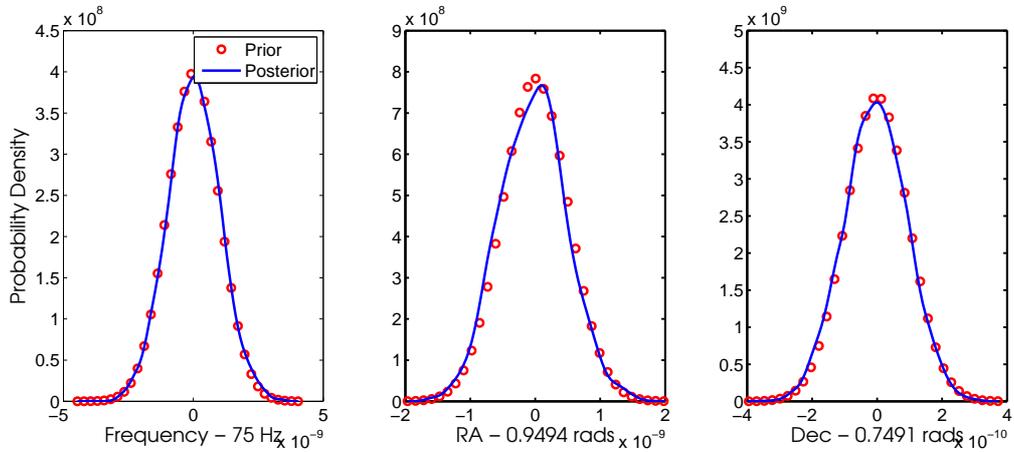}
\caption{The marginalised posteriors for frequency, right ascension and
declination calculated from data consisting of Gaussian noise, using the priors
shown.}\label{fig:priors}
\end{center}
\end{figure}

As in \cite{Abbott:2010} radio pulsar observations will often come with
uncertainties on the parameters, which although in general are very small can
mean that the single phase model used in \eref{eq:complexsignal} is not
sufficient to described the signal. These extra parameters describing the phase
change need to be incorporated into the \gw search. The new code can incorporate
these observed uncertainties, via the phase parameter covariance matrix, to be
used as priors. Figure~\ref{fig:priors} shows the marginalised posteriors for
three parameters (frequency, right ascension and declination) that have been
estimation from data just containing Gaussian noise, using priors similar to
uncertainties from radio observations, with other parameters held fixed. The
priors used are also plotted and show that for these phase parameters the
posteriors are completely defined by the priors as would be expected for data
containing only noise. These priors were also correlated, with the prior
correlation coefficients used and correlation coefficients calculated from
the posterior given in Table~\ref{tab:priors}. Again these show that the
expected prior distribution is being obtained. 
\begin{table}
\begin{tabular}{cc}

\begin{tabular}{c|ccc}
~ & RA & Dec & Frequency \\
\hline
RA & 1.0 & 0.5 & -0.2 \\
Dec & 0.5 & 1.0 & -0.7 \\
Frequency & -0.2 & -0.7 & 1.0
\end{tabular}

&

\begin{tabular}{c|ccc}
~ & RA & Dec & Frequency \\
\hline
RA & 1.0 & 0.467 & -0.173 \\
Dec & 0.467 & 1.0 & -0.694 \\
Frequency & -0.173 & -0.694 & 1.0
\end{tabular}

\end{tabular}
\caption{The correlation coefficients of the priors (left) on frequency, right
ascension and declination used to produce a posterior (right) for data
consisting of Gaussian noise.}\label{tab:priors}
\end{table}

\begin{figure}[!htbp]
\begin{center}
\includegraphics[width=0.7\textwidth]{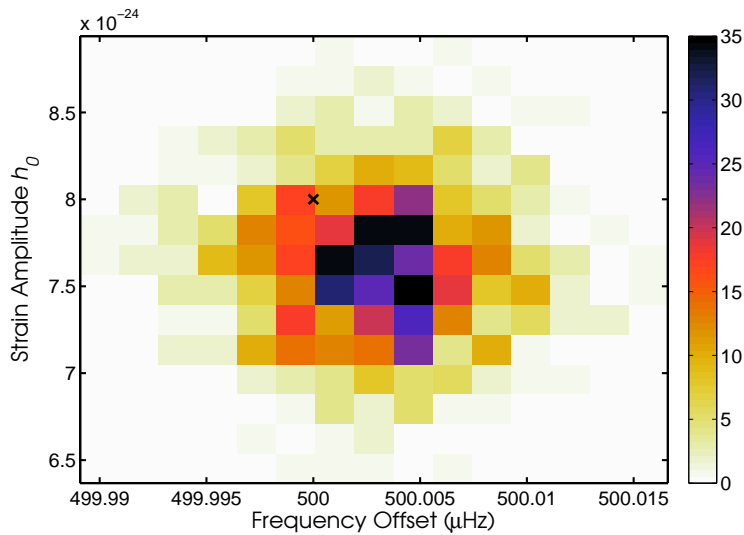}
\caption{The 2D posterior of signal frequency offset (from the central search
frequency) against $h_0$ for a fake pulsar signal with a signal-to-noise ratio
of $\sim 19$. The true injection parameters are marked with a black cross.
The prior volume covered $\pm10^{-3}$\,Hz in frequency offset and 0 to
$10^{-22}$ in $h_0$.}\label{fig:h0_vs_f0}
\end{center}
\end{figure}

There are also potential sources and candidates from all sky searches for
unknown pulsars/neutron stars (e.g.~\cite{Abbott:2009, Abadie:2011b}) that may
have parameters e.g.\ frequency, frequency derivatives and sky position, with
relatively large uncertainties spanning many phase models. The new code has
preliminarily been tested in a search search containing a fake signal, with a
prior range on the frequency band of $2\times 10^{-3}$\,Hz, with the recovered
$h_0$--frequency posterior shown in Figure~\ref{fig:h0_vs_f0}. The code recovers
the injected signal well, but in a future paper we will fully characterise the
code for searches over these wider frequency bands, and compare it to similar
searches using the $\mathcal{F}$-statistic as applied in \cite{Abbott:2008}, or
the $\mathcal{B}$-statistic of \cite{Prix:2009}.

\subsection{Model selection}

Nested sampling produces a value of the evidence which can be used for
model comparisons. In a future paper the code will be fully characterised for
comparing the evidence for the signal model described in Section~\ref{sec:intro}
against a model that the data contains just Gaussian noise. The ratio of these
evidences can be used as a detection statistic to assess a candidate signal's
significance. The code is also flexible enough to allow additional models
including extra parameters. We plan to assess a new signal model, which can
include emission at once and twice the rotation frequency \cite{Jones:2010} and
includes additional angular parameters, and compare it to the standard model
given by \eref{eq:complexsignal}. Further signal and noise models (for example
models including interference lines that are incoherent between multiple
detectors \cite{Keitel:2012}) could easily be added in the future in the current
framework.

\section{Conclusions}

A new code for parameter estimation and model comparison, based on the nested
sampling algorithm, has been developed for continuous-wave searches in \gw data.
It has been shown to be consistent with a previous code, used successfully in
many searches, in extracting signal parameters and producing upper limits on \gw
amplitude. The new code has advantages over the old code in that it requires
less tuning for specific searches and allows comparison between different
models. 

The new code is flexible enough to be expanded to larger numbers of parameters
and parameter ranges. This will allow it to be used in searches for pulsars
with less well defined parameters, or as a method to follow up candidates from
all-sky neutron star searches. These will be characterised in future papers. 
Additional MCMC proposal distributions for drawing new samples will also be 
investigated to study improvements in the algorithm's efficiency. These will 
including proposals that only update one parameter at a time when drawing from
a multivariate Gaussian, using a $k$-D tree of the samples to form a proposal
(e.g.\ \cite{Farr:2011}) and potentially using {\sc MultiNest}
\cite{Feroz:2009}, which has been integrated into the {\tt lalinference}
package of {\tt lalsuite} \cite{lalsuite}.

In the near future this code will be used in searches for a large number of
known pulsars in data from the recent LIGO S6 and the Virgo VSR2, 3 and 4 data
runs.

\ack
The authors would like to thank the LIGO Scientific Collaboration and Virgo
Collaboration continuous wave and CBC Bayesian parameter estimation working
groups, and the developers of the {\tt lalinference} software suite, for many
useful discussions. We also thank the referee for their many useful comments on
the draft. The document has LIGO DCC number LIGO-P1100171. 

\section*{References}
\bibliographystyle{iopart-num}
\bibliography{amaldi_proceedings}

\end{document}